# L1$_0$-FePd Synthetic Antiferromagnet Through a Face-centered-cubic Ruthenium Spacer Utilized for Perpendicular Magnetic Tunnel Junctions

De-Lin Zhang,[1] Congli Sun,[3] Yang Lv,[1] Karl B. Schliep,[2] Zhengyang Zhao,[1] Jun-Yang Chen,[1] Paul M. Voyles,[3] and Jian-Ping Wang[1*]

[1]Department of Electrical and Computer Engineering, University of Minnesota, 200 Union Street SE, Minneapolis, Minnesota 55455, USA

[2]Department of Chemical Engineering and Materials Science, University of Minnesota, 421 Washington Avenue SE, Minneapolis, Minnesota 55455, USA

[3]Department of Materials Science and Engineering, University of Wisconsin-Madison, Madison, WI 53706, USA

Magnetic materials that possess large bulk perpendicular magnetic anisotropy (PMA) are essential for the development of magnetic tunnel junctions (MTJs) used in future spintronic memory and logic devices. The addition of an antiferromagnetic layer to these MTJs was recently predicted to facilitate ultrafast magnetization switching. Here, we report a demonstration of bulk perpendicular synthetic antiferromagnetic (p-SAF) structure comprised of an (001) textured FePd/Ru/FePd trilayer with a face-centered-cubic *(fcc)* phase Ruthenium spacer. The L1$_0$-FePd p-SAF structure shows a large bulk PMA ($K_u$~10.2 Merg/cm$^3$) and strong antiferromagnetic coupling (-$J_{iec}$~2.60 erg/cm$^2$). Full perpendicular magnetic tunnel junctions (p-MTJs) with L1$_0$-FePd p-SAF layer were then fabricated. Tunneling magnetoresistance ratios of up to ~25% (~60%) are observed at room temperature (5 K) after post-annealing at 350 $^o$C. Exhibiting high thermal stabilities and large $K_u$, the bulk p-MTJs with a L1$_0$-FePd p-SAF layer could pave a way for next-generation ultra-high-density and ultralow-energy spintronic applications.

[*]Author to whom correspondence should be addressed: jpwang@umn.edu

## I. INTRODUCTION

Magnetic materials that exhibit perpendicular magnetic anisotropy (PMA) are promising candidates for the development of ultrahigh-density and ultralow-energy spintronic memory and logic devices, due to their high thermal stability, and scalability [1-4]. Recently, interfacial PMA materials have made considerable progress in the application of spin-transfer-torque magnetic random access memory (STT-MRAM) [5-8]. However, because of their relatively low PMA value ($K_u \sim$2-5 Merg/cm$^3$) and relatively large damping constant ($\alpha \sim$0.015-0.027) [5,9,10], they may not fully satisfy the scaling demands needed for next-generation spintronic memory and logic devices. When scaling spintronic devices to commercially sustainable sizes, like 10 nm nodes, large $K_u$ and low $\alpha$ values are required to realize longer retention times and ultra-low switching current densities. The manganese (Mn)-based Heusler alloys [11-15] and the L1$_0$-FePd are promising candidates for satisfying these requirements. Compared with the Mn-based Heusler alloys, the L1$_0$-FePd bulk PMA material possesses very attractive properties, such as a large $K_u$ (13-14 Merg/cm$^3$) [16,17], a low $\alpha$ (0.002) [18,19], and a low processing temperature (200 ºC) [20], which are summarized in Table 1. Furthermore, the switching current density ($J_c$) for spintronic memory devices, such as STT-MRAM, is a critical parameter, defined by the equation: $J_c = 2\alpha e t_F M_s \left( H_{appl} + H_k \right) / \hbar \eta$ [21], where $J_c$ mainly relates to the damping constant ($\alpha$), the saturation magnetization ($M_S$), and the perpendicular anisotropy field ($H_k$). From this equation it is clear that a small $M_S$ and low $\alpha$ value should be pursued to reduce the $J_c$ value. For L1$_0$-FePd thin films, the low $\alpha \sim$0.002 has been demonstrated experimentally, however, the $M_S$ of ~1100 emu/cm$^3$ is relatively high. A promising solution to lower its Ms is to develop a synthetic antiferromagnetic (SAF) structure [22,23], in which two ferromagnetic layers are coupled antiferromagnetically through a spacer so that the magnetization can be reduced. The SAF structure is also being pursued because it was theoretically predicted to significantly increase the switching speed and reduce the $J_c$ value in the MTJ devices [24,25]. Additionally, the velocity of domain wall motion in the SAF layers was found to be largely enhanced [26,27]. Current reports on the perpendicular SAF (p-SAF) structures have been increasingly focused on the [Co/Pd]$_n$ [28,29] and [Co/Pt]$_n$ [30,31] multilayer systems. However, the disadvantages of these p-SAF systems are that they are (111) texture, making it difficult to epitaxially grow on the MgO (001) tunnel barrier, have larger α values, and have limited $K_u$ values compared with the L1$_0$-FePd p-SAF structure investigated here.

In this work, we demonstrate for the first time a L1$_0$-FePd p-SAF structure and a L1$_0$-FePd synthetic antiferromagnetic perpendicular magnetic tunnel junction (named L1$_0$-FePd SAF p-MTJ). The L1$_0$-FePd p-SAF structure grown here with a (001) texture possesses a high $K_u$~10.2 Merg/cm$^3$ and low net remanent magnetization (~500 emu/cm$^3$). One of the most important discoveries here is the epitaxial growth of Ruthenium (Ru) spacer with a face-centered-cubic (*fcc*) phase on the L1$_0$-FePd thin film, which resulted in a large interlayer exchange coupling (IEC) $-J_{iec}$~2.60 erg/cm$^2$. This value is about one order of magnitude larger than that of the [Co/Pd]$_n$ or [Co/Pt]$_n$ p-SAF structures. Moreover, a tunnelling magnetoresistance (TMR) ratio of ~25.0% tested at room temperature (RT) was obtained in the $L1_0$-FePd SAF p-MTJ devices with the $L1_0$-FePd p-SAF layer after post-annealing at 350 °C. Furthermore, a TMR ratio of 13% is retained when the post-annealing temperature is increased up to 400 °C, implying that this kind of the FePd SAF p-MTJs can be integrated into the semiconductor process.

## II. EXPERIMENT

All the samples were deposited on (001) single crystal MgO substrates using ultra-high vacuum magnetron sputtering systems with base pressure less than $5.0 \times 10^{-8}$ Torr. During the deposition of the Cr(15 nm)/Pt(5 nm) seed layer, FePd, and Ru layers, the substrate temperature was kept at 350 °C. The pressure of Ar working gas was set at 4.5 mTorr for the FePd layer and 2.0 mTorr for the other layers. The FePd thin films were prepared by co-sputtering with the Fe and Pd targets. The composition was determined to be $Fe_{53.2}Pd_{46.8}$ by Rutherford backscattering spectrometry (RBS). After the FePd layer deposition, the films were cooled to RT. Subsequently, a 5-nm-thick Ta capping layer was grown on the single layer FePd film and FePd p-SAF structure to facilitate the investigation of their magnetic properties. The Ta(0.8)/CoFeB(1.3)/MgO(2)/CoFeB(1.3)/Ta(0.7)/[Pd(0.7)/Co(0.3)]$_4$/Ta(5) stack (the unit in nanometer) was grown on the FePd p-SAF structure to fabricate the full FePd SAF p-MTJ stack. The structural properties of the FePd p-SAF structure and the FePd SAF p-MTJ stack were characterized by out-of-plane (θ-2θ scan) x-ray diffraction (XRD) with Cu Kα radiation (λ=0.15418 nm) using a Bruker D8 Discover system and scanning transmission electron microscopy (STEM), whereas the magnetic properties were determined using a Physical Property Measurement System (PPMS). Cross-sectional STEM samples of the $L1_0$-FePd p-SAF structure and the $L1_0$-FePd SAF p-MTJ stack were prepared by an in-situ lift out method using a Zeiss Auriga focused ion beam (FIB) system. STEM imaging was performed on an FEI Titan with CEOS probe aberration corrector operated at 200 kV with a probe convergence angle of 24.5 mrad, spatial resolution of 0.08 nm, and probe current of ~20 pA.

The $L1_0$-FePd SAF p-MTJ stacks were patterned into micron-sized pillars with diameters ranging from 4~20 μm by a conventional photolithography and an Ar ion milling process. Electrical contacts were formed from Ti (10 nm)/Au (120 nm). The patterned MTJ devices were annealed from 300 $^o$C to 400 $^o$C by a rapid thermal annealing (RTA) process. Magnetotransport properties of the patterned MTJs were measured at different temperatures between 5 K and 300 K using a dc four-probe method using with a Quantum Design Physical Properties Measurement System (PPMS) with the Keithley 6221 current source and 2182 voltmeter. During the measurement, a magnetic field was applied along the out-of-plane (perpendicular) direction and the positive current was defined as the electron flow from the top reference layer to the bottom free layer of the $L1_0$-FePd SAF p-MTJ devices.

## III. RESULTS AND DISCUSSION

### A. Magnetic properties of the $L1_0$-FePd p-SAF structure

First we prepared an 8-nm-thick FePd single layer thin film on the (001) single crystalline MgO substrate to characterize the crystalline structure and PMA properties. The Cr (15 nm)/Pt (4 nm) buffer layer with *in-situ* substrate temperature of 350 $^o$C was used to induce the (001) texture. The $M_S$ was determined to be ~1050 emu/cm$^3$ from the magnetic hysteresis (M-H) loop, and $K_u$ was evaluated to be 11.5 Merg/cm$^3$ following the equation $K_u = \frac{M_S H_K}{2} + 2\pi M_S^2$ (1) [17], here the $H_K$ and $M_S$ are the saturation magnetic field and the saturation magnetization, respectively. These values are close to the values of the L1$_0$-FePd bulk material (~1100 emu/cm$^3$ and 18 Merg/cm$^3$) [16] (Fig. 1 in ref 32). After that the FePd p-SAF structures with a stack of FePd (3 nm)/Ru ($t_{Ru}$ nm)/FePd (3 nm) were grown using the same process as the FePd single layer thin film. The schematic of the FePd SAF structure is shown in the inset of Fig. 1(a), where two FePd PMA layers are coupled antiferromagnetically via a thin Ru spacer. The thickness of the Ru spacer ($t_{Ru}$) is varied from 0.9 nm to 1.4 nm to track the IEC between two FePd PMA layers. The observed largest IEC occurred when the thickness of the Ru spacer was around 1.1 nm. Its M-H loops are plotted in Fig. 1a. We find that the L1$_0$-FePd p-SAF structure presents good PMA with a square shape minor M-H loop and a net remanent magnetization of ~500 emu/cm$^3$. The $M_S$ of the L1$_0$-FePd p-SAF structure calculated from the M-H loops is ~960 emu/cm$^3$, which is a little lower than that of the single layer FePd thin film. One probable reason is the formation of the FePdRu$_x$ alloy at the Ru/FePd interface due to the Ru diffusion [33]. Meanwhile, the $H_K$ is determined to be ~8.9 kOe from the in-plane M-H loop. The PMA constant $K_u$ of the FePd p-SAF structure is then evaluated to be ~10.2 Merg/cm$^3$ following the equation used before, which is several times larger than that of interfacial PMA materials (e.g. Ta/CoFeB/MgO structure).

Normally, the shape of the out-of-plane M-H loop of the p-SFA systems depends on the competition between the PMA and the antiferromagnetic coupling, so one can determine the type of the antiferromagnetic coupling exhibited from the M-H loop of the p-SFA sample [34]. For the FePd p-SAF structure as shown in Fig. 1(a), we can clearly observe the spin-flop switching between two FePd layers at the high external magnetic field. Meanwhile, the spin-flip switching appears at the low external magnetic field, implying that two FePd layers form the antiferromagnetic alignment. This also indicates that the strength of antiferromagnetic coupling between two FePd layers is larger than that of the PMA of FePd SAF structure. Following the equation $-4J_{iec} = H_{ex}M_S t_{FM} + 2K_{u,eff} t_{FM}$ (in this case, $K_u < -J_{iec}/t$) [34], the $J_{iec}$ of the FePd p-SAF structure was calculated to be ~-2.60 erg/cm$^2$, where $K_{u,eff}$ is the magnetic anisotropy, $H_{ex}$ is the exchange field (~9.2 kOe for FePd SAF structure) and $t_{FM}$ is the thickness of the FePd PMA layer. This value is about one order of magnitude larger than that of the [Co/Pd]$_n$ p-SAF system with the same post-annealing temperature[29]. In addition, previous reports show that the moderate antiferromagnetic exchange coupling performance can be observed at the second peak of oscillation with $t_{Ru}$~0.9 nm [35]. The difference in the optimal Ru thickness ($t_{Ru}$~1.1 nm) in this work is attributed to the relatively large roughness (~0.30 nm) from the Cr/Pt seed layer and the high deposition temperature used. This result is similar to that of the post-annealed [Co/Pd]$_n$ p-SAF structure with the strongest $J_{iec}$ when the $t_{Ru}$ is ~1.3 nm thick [29].

**B. Structural properties of the L1$_0$-FePd p-SAF structure**

The crystalline structure of the L1$_0$-FePd p-SAF structure with a 1.1 nm thick Ru spacer was characterized by X-ray diffraction (XRD). The results are shown in Fig.

1(b). The (001) and (002) peaks from the $L1_0$-FePd layers and (002) peak from Cr and Pt seed layers are visible clearly, suggesting a well-formed superlattice structure. Because the $t_{Ru}$ ~1.1 nm is thin, it is difficult to observe the Ru peak by the XRD measurement. To identify the crystalline structure of the Ru spacer and the epitaxial relationship of the Ru and FePd layers, aberration corrected scanning transmission electron microscopy (STEM) was performed to characterize the atomic structures. Figure 2(a) shows the high angle annular dark field (HAADF) STEM images of the $L1_0$-FePd p-SAF structure (HAADF signal scales with $Z^a$, here Z is the atomic number of element and the "a" is the coefficient, so the image is dominated by high-Z atomic sites). As shown in Fig. 2(a), the (001) epitaxial relationship was observed throughout the film, starting with the MgO (001) surface and continuing through each layer of the stack along the vertical direction. Meanwhile, it is found that the Cr layer was oxidized in the Cr/MgO interface, and the Pt/Cr interface is rough but other layers form the sharp interface. The crystalline structure of the Ru spacer and the FePd layer is determined by the magnified HAADF-STEM image as shown in Fig. 2(b). From the STEM image, one can see that the bottom FePd layer matches the (001) texture of Pt/Cr seed layer with the epitaxial relationship Cr[110](001)||Pt[100](001)||FePd[100](001). The average lattice spacing of two atomic planes along (001) direction of the FePd layer measured from STEM image is 0.187 nm close to the lattice constant of the tetragonal AuCu-type FePd structure shown in Fig. 2(c).

Normally, the Ru spacer forms the hexagonal-close-packed (*hcp*) phase like in the Co/Ru [36] and [Co/Pd]$_n$/Ru SAF systems [29], generating the periodic oscillations of antiferromagnetic coupling due to the Ruderman-Kittel-Kasuya-Yosida (RKKY) interaction [36]. In our experiment, The Ru spacer interestingly also follows the (001)-orientated growth of the bottom FePd layer. The average in-plane lattice spacing is estimated to be 0.192 nm from the STEM image, which matches very well with the lattice spacing of the face-centered-cubic (*fcc*) phase Ru structure predicted by first principles calculation [35], as shown in Fig. 2(d). The highly textured growth of the Ru spacer on the FePd layer is due to a very small lattice mismatch (0.62%) between the L1$_0$-FePd and the *fcc*-phase Ru layers. This also induces the (001) texture of the top FePd layer with L1$_0$-phase. The epitaxial relationship of the FePd/Ru/FePd trilayer was also determined to be FePd[100](001)||Ru[100](001)||FePd [100](001). In addition, the STEM image with a 90-degree in-plane rotation of the TEM sample was used to confirm the texture of the FePd/Ru/FePd stack. The same epitaxial relationship as shown in Fig. 2(a) was observed in the FePd/Ru/FePd trilayer (Fig. 2 in ref 32). Those results of the L1$_0$-FePd p-SAF structure indicate that not only the Ru spacer on the (001) texture FePd layer presents the *fcc*-phase but also the *fcc*-phase Ru spacer can result in a larger antiferromagnetic coupling.

### C. Magnetic properties of the L1$_0$-FePd SAF p-MTJ stack

Based on the developed L1$_0$-FePd p-SAF structure, which has a relatively smooth surface with root-mean-square (RMS) surface roughness of ~0.30 nm (Fig. 3 in ref 32), we designed and fabricated a full L1$_0$-FePd SAF p-MTJ stack as shown in Fig. 3(a). Using Ta or other metal layer to couple a hard magnetic layer with CoFeB has been proposed and established as a standard for STT-RAM cells [37]. In this L1$_0$-FePd SAF p-MTJ stack, a composite layer with a stack of FePd/Ru/FePd/Ta/CoFeB was designed as the bottom free layer, where the FePd p-SAF trilayer couples with the CoFeB layer through a thin Ta layer. Meanwhile, a composite layer with a stack of CoFeB/Ta/[Co/Pd]$_n$ was designed as the top reference layer. A 1.3-nm-thick Co$_{20}$Fe$_{60}$B$_{20}$ (CoFeB) layer is introduced adjacent the MgO barrier to enhance the MR ratio. An ultra-thin Ta layer (~0.7 nm) is inserted between CoFeB and FePd ([Co/Pd]$_n$) layers to mitigate Pd diffusion that occurs during the high temperature annealing process [38].

The quality of the MgO tunnel barrier plays a significant role in obtaining a large TMR ratio in the MgO-barrier MTJ devices. The annular bright field (ABF) STEM was employed to determine the atomic structure of MgO barrier in the L1$_0$-FePd SAF p-MTJ stack post-annealed at 350 °C. From the ABF-STEM image shown in Fig. 3b, we can see that the MgO tunnel barrier layer crystallizes into a (001) textured structure, and a sharp MgO/CoFeB interface can be observed. The magnetic properties of the FePd p-SAF free layer and the L1$_0$-FePd SAF p-MTJ stack are characterized individually after the samples were post-annealed at 350 °C. Their out-of-plane M-H loops are shown in Fig. 3(c) and 3(d), respectively. From the M-H loop shown in Fig. 3(c), it is evident that the FePd p-SAF free layer possesses PMA with an antiferromagnetic coupling between the FePd p-SAF and CoFeB layers. The H$_C$ of the FePd p-SAF free layer is found to be ~390 Oe. The saturation magnetic field H$_s$ of the FePd free layer is slightly enhanced compared with the L1$_0$-FePd p-SAF stack. This $H_s$ enhancement is similar to what is observed in the CoFeB/Ta/CoFeB p-SAF stack [38].

Before integrating the $[Co/Pd]_n$ reference layer into the $L1_0$-FePd SAF p-MTJ stack, its magnetic property and thermal stability were studied. After post-annealing using the same experimental condition as the FePd free layer, the $[Co/Pd]_n$ reference layer showed a square shape M-H loop with Hc ~1400 Oe, and a high quality (111) texture was also observed (Fig. 4 in ref 32). Figure 3d depicts the magnetic properties of the $L1_0$-FePd SAF p-MTJ stack with post-annealing temperature of 350 °C. Three-step switching of magnetization was observed. The first and second switching fields, $H_{sw1}$~39 Oe and $H_{sw2}$~390 Oe, correspond to the CoFeB and $L1_0$-FePd p-SAF layers, respectively. The third switching is from the $[Co/Pd]_n$ reference layer with $H_{sw}$~700 Oe, which is smaller than that of the $[Co/Pd]_n$ reference layer on Si/SiO$_2$ substrate. The main reason is that the (001) texture of the bottom $L1_0$-FePd p-SAF free layer results in the lattice mismatch and affects the (111) texture of the $[Co/Pd]_n$ layer. The M-H loops of the FePd free layer and the $L1_0$-FePd SAF p-MTJ stack post-annealed at 400 °C show the same trend as the sample post-annealed at 350 °C (Fig. 5 in ref 32).

## D. Magnetotransport properties of the $L1_0$-FePd SAF p-MTJ devices

To study the magnetotransport property and thermal stability of the proposed structures, the $L1_0$-FePd SAF p-MTJ stack illustrated in Fig. 3(a) was patterned into micron-size pillars with diameters ranging from 4-20 μm using a standard lithography patterning process. After patterning, these devices were post-annealed from 300 °C to 400 °C by a rapid thermal annealing (RTA) process. The TMR versus external magnetic field (TMR-H) loops of the $L1_0$-FePd SAF p-MTJ devices presented in the inset of Fig. 4(e) were tested at RT by a standard four-probe resistance measurement technique. The TMR-H loops of the 12-μm-diameter $L1_0$-FePd SAF p-MTJ devices are plotted in Figs. 4(a)~4(d). From these RT TMR-H loops, we can observe a plateau in the high-resistance state and a sharp magnetization switching between the parallel state and the antiparallel state while sweeping the perpendicular external magnetic field. This suggests that the FePd free layer and the $[Co/Pd]_n$ reference layer possess a good PMA even after post-annealing up to 400 °C. As depicted in Fig. 4(e), with the increase of the post-annealing temperature from 300 °C to 350 °C, the TMR ratio slightly increases from ~24% to ~25%. Then the TMR ratio decreases to ~13% with the increasing of the annealing temperature up to 400 °C, which indicates the FePd SAF p-MTJs devices have a relatively good thermal stability. The decrease of TMR ratio for the $L1_0$-FePd SAF p-MTJ devices may be due to three reasons: one reason is the interlayer diffusion of the Ta or Pd atoms into the CoFeB layer to form the $CoFeBTa_x$ or $CoFeBPd_x$ thin layers. Another reason is that the CoFeB layer is oxidized at the CoFeB/MgO interface to form the dead layer. The third reason is that Boron (B) atoms diffuse into the MgO tunnel barrier after the high-temperature thermal treatment. These reasons lead to low spin-polarization of the CoFeB layer, create surface magnetization instability, and produce elemental impurities in the MgO tunnel barrier. These factors can strongly affect the magnetotransport properties of the p-MTJs [40,41]. For the MgO-barrier MTJs, the giant TMR ratio is attributed to the spin-dependent coherent tunnelling through the $\Delta_1$ Bloch state in high quality (001) epitaxial MgO tunnel barrier [42].

In order to understand the tunnelling behaviour (coherent or incoherent tunnelling) of the $L1_0$-FePd SAF p-MTJ devices, we investigate the TMR ratio as a function of the temperatures. The MR-H loops were tested from 5 K to 300 K for the 12-μm-diameter $L1_0$-FePd SAF p-MTJ devices post-annealed at 350 °C and 400 °C, respectively, by a Quantum Design Physical Properties Measurement System (PPMS). The results are plotted in Fig. 4(f) and Fig. 4(g). We can clearly see that the TMR ratio gradually increases from ~25% up to ~60% for the junctions annealed at 350 °C and from ~13% up to ~27% for the junctions annealed at 400 °C while the testing temperature goes down from 300 K to 5 K. The increase of TMR ratio mainly originates from the increase of the $R_{AP}$ value, which dominates the TMR ratio. However, the $R_P$ value presents weak temperature dependence for both p-MTJs, as shown in Figs. 4(h) and 4(i). Coherent tunnelling behaviour is illustrated by dramatic increase in $R_{AP}$ while $R_P$ remains constant when the testing temperature is decreased [42]. From the Figs. 4(h) and 4(i), we can deduce that the $L1_0$-FePd SAF p-MTJ devices annealed at 350 °C and 400 °C show non-perfect coherent tunnel behaviour. This explains why the $L1_0$-FePd SAF p-MTJ devices show the relatively low TMR ratio at RT, which may be due to the elemental diffusion or dead layer mentioned before.

To further understand these devices, the normalized TMR ratio as a function of bias voltage (V) was investigated. The $V_{half}$ defined as the bias voltage at which the TMR ratio drops to one-half of the zero-bias value is a crucial factor for the device application in ultra-high-density MRAM. The $V_{half}$ of the $L1_0$-FePd SAF p-MTJ devices annealed at 350 °C (400 °C) is determined to be ~380 (~370) mV and ~400 (~470) mV for positive and negative bias voltage. This is just a little lower than that of $Co_2FeAl$ (theoretical 100% spin polarization of the half-metallic Heusler alloy) in-plane MTJs with the MgO tunnelling barrier (500 mV for positive and 600 mV for negative bias directions) ( Fig. 6 in ref 32 ).

## IV. CONCLUSIONS

We have realized a novel bulk perpendicular SAF structure and the integration of p-MTJ stack using the $L1_0$-PMA FePd thin films. The (001) epitaxial $L1_0$-FePd p-SAF structure shows large antiferromagnetic coupling ($-J_{iec}$~2.60 erg/cm$^2$) with a high $K_u$~10.2 Merg/cm$^3$ and a low net remanent magnetization (~500 emu/cm$^3$). High RT TMR ratios up to ~25% (~13%) were achieved for the $L1_0$-FePd SAF p-MTJ devices post-annealed at 350 °C (400 °C) using the FePd free layer, which suggests that the $L1_0$-FePd SAF p-MTJs could adhere to standard back-end-of-line (BEOL) processes. Furthermore, this demonstrated $L1_0$-FePd p-SAF structure could be used to study the domain wall motion, spin-orbit torques [43], skyrmion [44] and antiferromagnetic spintronics [45]. These combined results provide significant potential in scaling p-MTJs below 10 nm for applications on spintronic memory and logic devices. Further optimization of the deposition process and patterning process is under way to improve its MR ratio.

## ACKNOWLEDGMENTS

This work was supported by C-SPIN, one of six centers of STAR net, a Semiconductor Research Corporation program, sponsored by MARCO and DARPA, and by NSF (ECCS-1310338). Jian-Ping Wang also thanks the Robert F Hartmann Endowed Chair Professorship of electrical engineering. We would like to thank Mr. Timothy Peterson for their help on the usage of PPMS and Mr. Patrick Quarterman and Prof. Bin Ma from University of Minnesota for useful discussion.

TABLE 1 Comparison of magnetic properties and TMR ratio of the PMA Mn-based Heusler films and the L10-phase PMA FePd film.

|  | **Mn-based Heusler** | **$L1_0$-phase FePd** |
|---|---|---|
| **Spin polarization** | 58% [11] | --- |
| **Damping constant** | 0.015~0.008[12]; 0.03[13] | 0.007[17]; 0.002~0.004[19] |
| **Magnetization** | 150 ~500 emu/cc | <500 emu/cc (p-SAF) |
| **Magnetic anisotropy** | <10 Merg/cc (t< 20 nm) | 11 Merg/cc (3.5 nm)[17] |
| **Thickness (t) in MTJ** | > 20 nm | 3~7 nm |
| **Lattice constant** | a=3.92 Å, c= 7.10 Å | a=3.90 Å, c=3.72 Å |
| **RT TMR ratio** | 24% (300 °C)[14] | 27% (325 °C)[15]; 25% (350 °C) (this work) |

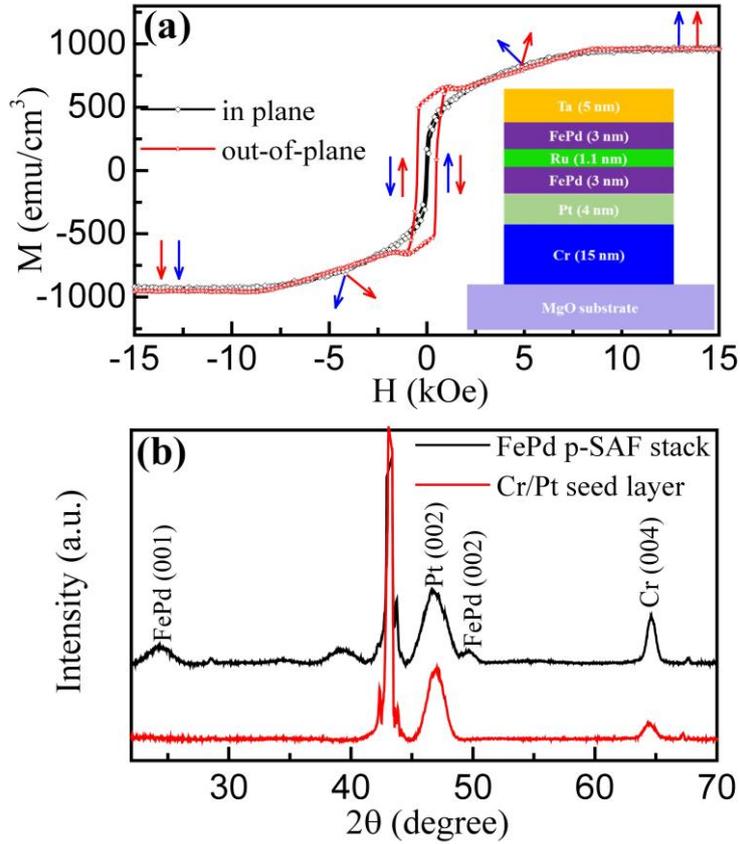

**FIG. 1.** (a) The magnetic hysteresis (M-H) loops of the FePd (3 nm)/Ru (1.1 nm)/FePd (3 nm) perpendicular synthetic antiferromagnetic (p-SAF) structure measured at room temperature. The inset of Fig. 1a shows the schematic diagram of the $L1_0$-FePd p-SAF structure. The Cr and Pt buffer layers are used to induce the (001) texture and the thin Ru layer is used as a spacer. (b) XRD pattern with out-of-plane θ-2θ scans for the $L1_0$-FePd p-SAF structure, the Cr/Pt buffer layer is used as a reference.

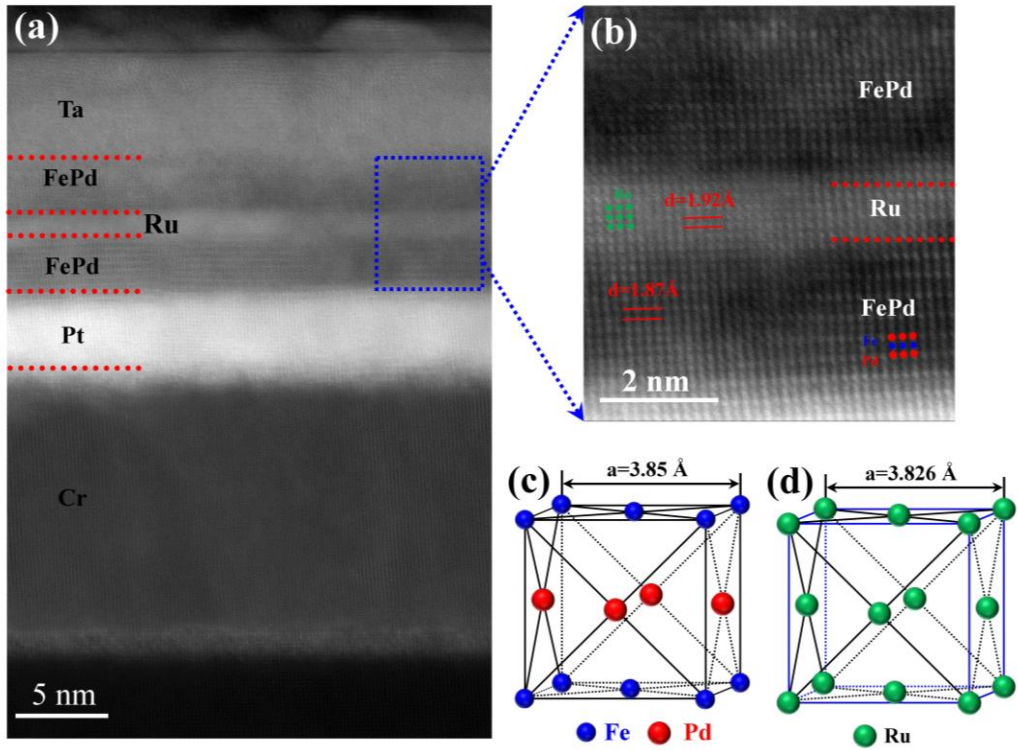

**FIG. 2.** (a) A high angle annular dark field (HAADF) STEM images for the $L1_0$-FePd p-SAF structure, the clearly epitaxial layered structure is observed. (b) A magnified STEM image of FePd/Ru/FePd trilayer, which shows the smooth interface between the FePd and Ru layers. Meanwhile, the lattice spacing of the FePd and Ru layers was calculated from the TEM, labelling in TEM image. (c) The crystalline structure of the FePd material with the tetragonal AuCu-type phase. The blue (small) ball denotes the Fe atom and red (big) ball represents the Pd atom. (d) The crystalline structure of Ru material with face-centered cubic (*fcc*) phase. The *fcc* phase Ru thin films has not been experimental realized just theoretically predicted by first-principle calculation. Its theoretical lattice constant is a=b=c=0.3826 nm.

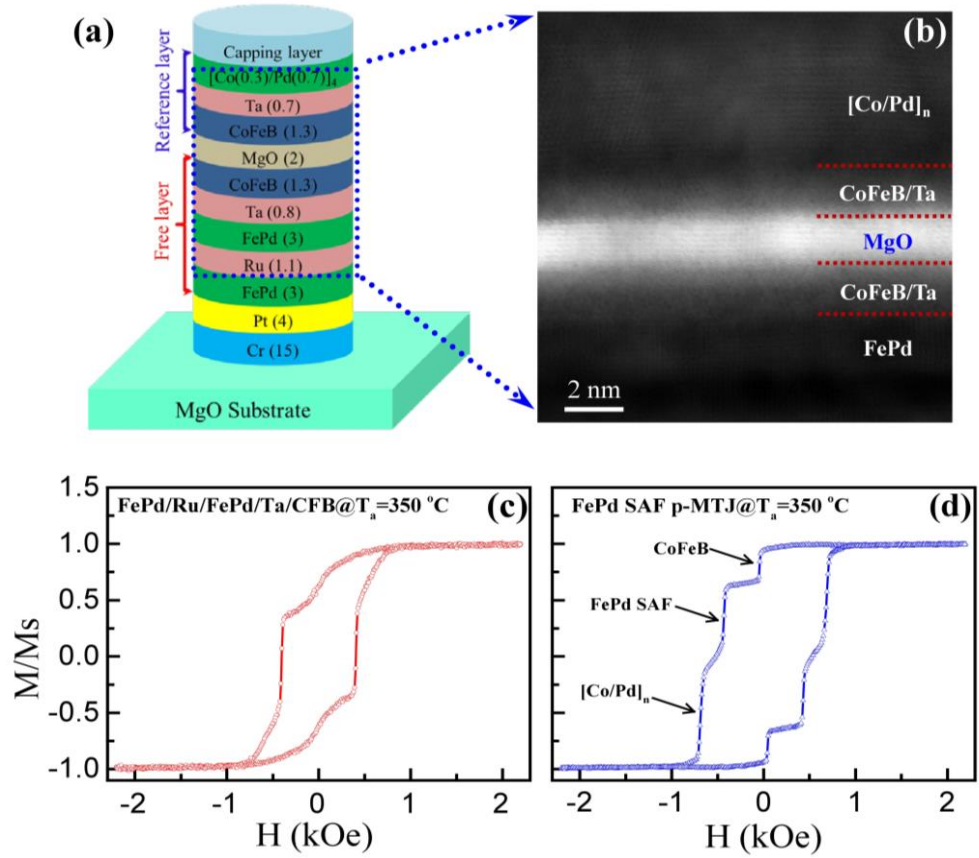

**FIG. 3.** (a) A schematic illustration of the $L1_0$-FePd SAF p-MTJ stack, in which the $L1_0$-FePd p-SAF layer couples with CoFeB layer through a thin Ta layer to form the bottom free layer. The $[Co/Pd]_n$ and CoFeB layers form the top reference layer by a thin Ta layer. The thin Ta layer plays the very important roles as coupling spacer and diffusion barrier which can be used to block the Pd diffusion when the MTJ devices are annealed at high temperature. (b) The ABF-STEM image of the CoFeB/MgO/CoFeB region of the $L1_0$-FePd SAF p-MTJ stack, which is used to determine the quality of MgO tunnelling barrier. (c, d) The room temperature out-of-plane M-H loops of the FePd free layer with a tack of FePd/Ru/FePd/Ta/CoFeB and the $L1_0$-FePd SAF p-MTJ stack. These stacks were post-annealing at 350 $^o$C by RTA process.

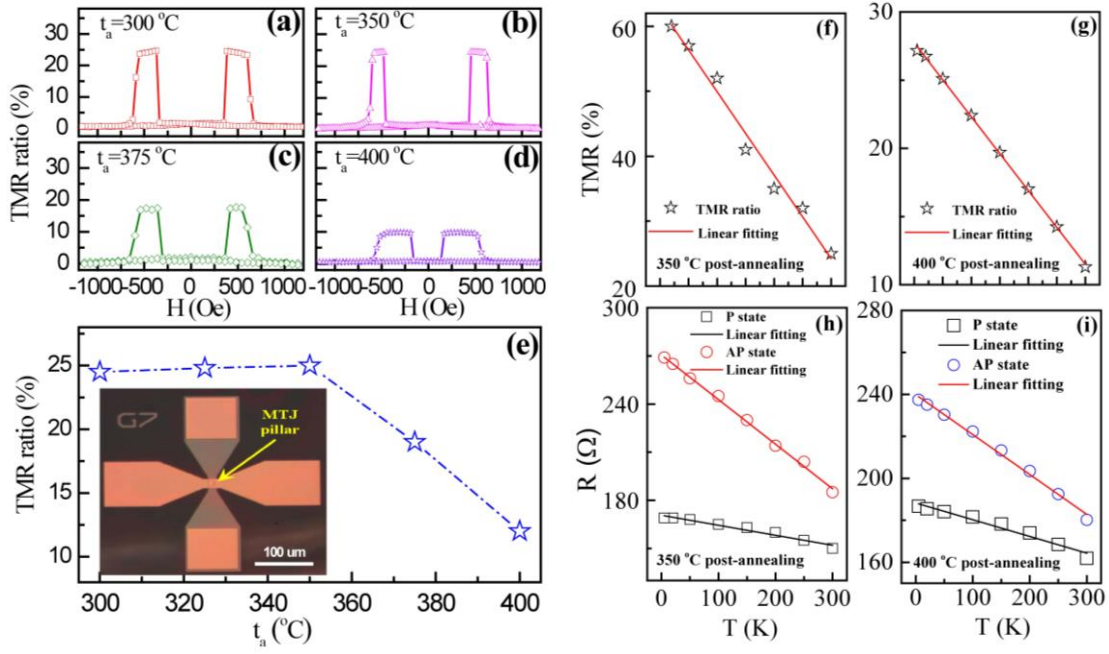

**FIG. 4.** The tunnelling magnetoresistance versa external magnetic field (MR-H) loops of the L1$_0$-FePd SAF p-MTJ devices post-annealed by RTA at (a) 300 $^o$C, (b) 350 $^o$C, (c) 375 $^o$C and (d) 400 $^o$C. The testing was carried out at room temperature. The external magnetic field is swapping from -1500 Oe to +1500 Oe along perpendicular plane of devices. (e) The TMR ratio as a function of the post-annealing temperatures of the L1$_0$-FePd SAF p-MTJ devices. The inset is the optical microscopy image of the real L1$_0$-FePd SAF p-MTJ device. (f) and (g) TMR ratio as a function of annealing temperatures for L1$_0$-FePd SAF p-MTJ devices annealed at 350 $^o$C and 400 $^o$C, respectively. (h) and (i) The temperature dependence on the resistance of the parallel state (open squares) and the antiparallel state (open circles) in the L1$_0$-FePd SAF p-MTJ devices annealed at 350 $^o$C and 400 $^o$C, respectively.